\documentclass[twocolumn]{emulateapj}
\usepackage{graphicx}
\usepackage{color}
\usepackage{hyperref}
\usepackage{amsmath}
\usepackage{color}
\usepackage[dvipsnames]{xcolor}

\def\arcsec{\hbox{$^{\prime\prime}$}}
\def\deg{\hbox{$^\circ$}}
\def\init{\hspace{0.75 mm}}
\def\gal{J1224+5555}

\renewcommand{\baselinestretch}{1.02}

\begin{document}

\title{Obscured AGNs in Bulgeless Hosts discovered by WISE: The Case Study of SDSS \gal}

\author{S.\init Satyapal\altaffilmark{1,7},  N.\init J.\init Secrest\altaffilmark{1,2}, B.\init Rothberg\altaffilmark{1,3},  J. O'Connor\altaffilmark{1}, S.\init L.\init Ellison\altaffilmark{4}, R. C. Hickox\altaffilmark{5}, A.\init Constantin\altaffilmark{6}\ ,M.\init Gliozzi\altaffilmark{1}, and  J.\init L.\init Rosenberg\altaffilmark{1}}

\altaffiltext{1}{George Mason University, Department of Physics \& Astronomy, MS 3F3, 4400 University Drive, Fairfax, VA 22030, USA}

\altaffiltext{2}{United States Naval Observatory, Washington, DC 20392, USA}

\altaffiltext{3}{LBT Observatory, University of Arizona, 933 N.~Cherry Ave., Tuscan, AZ 85721, USA}

\altaffiltext{4}{Department of Physics and Astronomy, University of Victoria, Victoria, BC V8P 1A1, Canada}

\altaffiltext{5}{Department of Physics and Astronomy, Dartmouth College, 6127 Wilder Laboratory, Hanover, NH 03755, USA}

\altaffiltext{6}{Department of Physics and Astronomy, James Madison University, PHCH, Harrisonburg, VA 22807, USA}

\altaffiltext{7}{Department of Astronomy, University of Maryland, College Park, MD 20742, USA}

\begin{abstract}
There is mounting evidence that  supermassive black holes form and grow in bulgeless galaxies.  However, a robust determination of the fraction of AGNs in bulgeless galaxies, an important constraint to  models of supermassive black hole seed formation and merger-free models of AGN fueling, is unknown, since optical studies have been shown to be incomplete for low mass AGNs. In a recent study using the \textit{Wide-field Infrared Survey Explorer}, we discovered hundreds of bulgeless galaxies that display mid-infrared signatures of extremely hot dust suggestive of powerful accreting massive black holes, despite having no signatures of black hole activity at optical wavelengths.  Here we report X-ray follow-up observations of J122434.66+555522.3, a  nearby ($z=0.052$) isolated bulgeless galaxy that contains an unresolved X-ray source detected at the 3$\sigma$ level by \textit{XMM-Newton} with an observed luminosity  uncorrected for intrinsic absorption of  $L_\mathrm{2-10~keV}=(1.1\pm0.4)\times10^{40}$~erg~s$^{-1}$. Ground-based near-infrared spectroscopy with the {\it  Large Binocular Telescope} together with multiwavelength observations from ultraviolet to millimeter wavelengths together  suggest that \gal\ harbors a highly absorbed AGN with an intrinsic absorption  of ~$N_\mathrm{H} >10^{24}$~cm$^{-2}$. The hard X-ray luminosity of the putative AGN corrected for absorption is $L_\mathrm{2-10~keV}\sim3\times10^{42}$~erg~s$^{-1}$, which, depending on the bolometric correction factor, corresponds to a bolometric luminosity of the AGN of   $6\times10^{43}$~erg~s$^{-1}\lesssim L_\textrm{bol.}~ \lesssim3\times10^{44}$~erg~s$^{-1}$, and a lower mass limit for the black hole of  $M_\textrm{BH}\simeq2\times10^6~M_\sun$, based on the Eddington limit. While enhanced X-ray emission and hot dust can be produced by star formation in extremely low metallicity environments typical in dwarf galaxies, \gal\ has a stellar mass of $\sim2.0\times10^{10}~M_\sun$, and an above solar metallicity (12+$\log{\textrm{O/H}}$=9.11), typical of our {\it WISE}-selected bulgeless galaxies sample. While collectively, these observations suggest the presence of an AGN, we caution that identifying obscured AGNs in the low luminosity regime is challenging, and often requires multi wavelength observations.These observations suggest that low luminosity AGNs can be heavily obscured and reside in optically quiescent galaxies, adding to the growing body of evidence that the fraction of  bulgeless galaxies with accreting black holes may be significantly underestimated based on optical studies. 

\end{abstract}


\section{Introduction}
\label{intro}

Over 40 supermassive black holes (SMBHs) with masses $\sim10^{9}~M_\sun$  have
been discovered at z $>$ 6, including one with $\sim10^{10}~M_\sun$  \citep[e.g.,][]{fan2006, Wu2015, Mortlock2011, Venemans2013}.  The existence of such massive black holes when the universe was less than 1 billion years poses significant challenges to theories of the origin of SMBH seeds and their subsequent growth. The seed population could have formed from the remnants of primordial stars \citep[e.g.,][]{volonteri2003, volonteri2005, johnson2007}, which would create the lowest mass seeds ($\sim10^{2}~M_\sun$), or in dense star clusters through runaway stellar mergers  \citep[e.g.,][]{ Begelman1978, Ebisuzaki2001}, which would create intermediate mass seeds ($\sim10^{4}~M_\sun$), or from the direct collapse of self-gravitating pre-galactic gas disks at high redshifts in cooling halos \citep[e.g.,][]{Haehnelt1993,Bromm2003,Agarwal2014} which would produce the most massive seeds ($\sim10^{4}-10^{6}~M_\sun$) (see review by Natarajan 2014). While the lowest mass seeds are the easiest to form, early continuous and extremely efficient growth of the seeds required for the production of massive quasars at z$>$6 poses significant challenges, a problem which is exacerbated by recent simulations of early star formation which shows that the progenitor population III stars may be even less massive than initially thought due to turbulence and efficient fragmentation which inhibits the growth of massive stars \citep[e.g.,][]{Alvarez2009,Greif2011,Regan2014,SafranekShrader2014}. Discriminating between these various seed formation mechanisms is challenging, since seed formation occurs at redshifts z $>$ 15, and is therefore observationally inaccessible.  However as pointed out by several recent studies   \citep[e.g.,][]{Volonteri2009, Volonteri2010, vanwassenhove2010, greene2012}, the occupation fraction and mass function of nuclear black holes in local low mass galaxies and galaxies that lack classical bulges can provide important constraints on the original seed population, since these galaxies have presumably had a quiescent cosmic history and therefore their black holes carry the imprint of the original seed population. Apart from providing constraints on models of the origins of SMBHs, the search for active galactic nuclei (AGNs) in the small bulge mass regime is of intrinsic interest, since it is required in order to obtain a complete census of AGN activity in the local universe. Furthermore, the  mass and accretion rates of these SMBHs provide insight into our understanding of the efficiency of secular processes in the growth of nuclear black holes \citep[e.g.,][]{hopkins2014}. As a result, the search for black holes in the low bulge mass regime has been an area of active research over the past decade. 

In recent years, there have been a number of discoveries of AGNs in low mass galaxies or galaxies that lack classical bulges, suggesting that black holes do exist in the low bulge mass regime and are more common than previously thought \citep[e.g.,][]{filippenko2003, barth2004, satyapal2007, greene2007a, satyapal2008, satyapal2009, dewangan2008, shields2008, ghosh2008, izotov2008, desroches2009, gliozzi2009,mcalpine2011, jiang2011, reines2011, reines2012, secrest2012, ho2012, secrest2013, simmons2013, dong2012, araya2012, reines2013, coelho2013, schramm2013, bizzocchi2014, satyapal2014, reines2014, maksym2014, moran2014, yuan2014,secrest2015,miller2015, sartori2015,mezcua2016,lemons2015}. However, despite the recent advent of the vast amount of optical spectroscopic data available from the Sloan Digital Sky Survey, to date, there still exist only small samples of AGNs discovered by optical surveys in small-bulge galaxies and only a handful with no evidence for any bulge. Indeed, a key and striking result based on optical spectroscopic studies, is that the fraction of galaxies with signs of accretion activity drops dramatically at stellar masses  $\log{M_\star/M_\sun}<10$~\citep[e.g.][]{kauffmann2003}. For a sample of dwarf galaxies with stellar masses $\log{M_\star/M_\sun}<9.5$ and high quality optical emission line measurements, only 0.1\% of galaxies are unambiguously identified as AGNs based on their optical line ratios \citep{reines2013}.

Using the all-sky \citep[\textit{Wide-Field Infrared Sky Explorer Survey; WISE};][]{wright2010}, we discovered several hundred optically normal bulgeless galaxies that display red mid-infrared colors [3.4$\micron$]-[4.6$\micron$] (hereafter $W1$-$W2$) suggestive of dominant AGNs  \citep[]{stern2012} that may outnumber optically-identified AGNs by as much as a factor of $\approx6$ \citep[]{satyapal2014}. Most of these galaxies do not show signatures of AGNs in their optical emission line ratios, suggesting heavy obscuration or dilution of emission line ratios by vigorous star formation, both of which are well-known limitations to optical spectroscopic surveys in finding low luminosity AGNs \citep[e.g.,][]{goulding2009,hopkins2009,trump2015}.

While the red mid-IR colors discovered by \textit{WISE} are highly suggestive of accretion activity, it is possible that the dust can be heated by star formation alone, and there are a number of blue compact dwarfs (BCDs) where this is indeed taking place \citep[e.g.,][]{griffith2011,izotov2014} . On the other hand, if the majority of bulgeless galaxies that display red mid-infrared colors are in fact AGNs, the fraction of AGNs in the low bulge mass regime has been significantly underestimated by optical studies. 

In a pilot study using \textit{XMM-Newton},  we obtained follow-up X-ray observations  of two bulgeless galaxies with red {\it WISE} colors in order to confirm the AGN.  In  \citet[]{secrest2015}, we presented the \textit{XMM-Newton} observation of the bulgeless galaxy SDSS~J132932.41+323417.0, a blue, irregular dwarf galaxy ($\log{M_\star /M_\sun}$=8.3) with a metallicity of $Z/Z_\sun$=0.4, at a redshift of $z$=0.0156, which  contains a hard, unresolved X-ray source detected by \textit{XMM-Newton} with luminosity  $L_\textrm{2-10~keV}=2.4\times10^{40}$~erg~s$^{-1}$, over two orders of magnitude greater than that expected from star formation, providing convincing evidence for the presence of an accreting massive ($10^4$-$10^5$~M$_\sun$) black hole.

In this paper, we present the  \textit{XMM-Newton} observations of the nearby ($z=0.052$) isolated bulgeless galaxy SDSS~J122434.66+555522.3 (hereafter J1224+5555) together with ground-based near-infrared spectra obtained with the\textit{ Large Binocular Telescope (LBT)}. Unlike SDSS~J132932.41+323417.0 (hereafter, J1329+3234),  \gal\ is not a dwarf galaxy.  It has a mass of  $\log{M_\star /M_\sun}$=10.3, comparable to the mean mass of the bulgeless sample presented in Table 1 of \citet{satyapal2014}, and a metallicity that is above solar (12+$\log{\textrm{O/H}}$=9.11), based on \citet{tremonti2004}. \gal\ hosts an unresolved mid-infrared nuclear source with extremely red colors typical of powerful AGNs ($W1$-$W2$=1.102~mag.; $W2$-$W3$=3.564~mag based on the final \textit{WISE} all-sky data release catalog (AllWISE\footnote{\url{wise2.ipac.caltech.edu/docs/release/allwise/}}). Based on the optical emission line ratios from the SDSS spectrum, there is no evidence for an AGN based on either the \citet{kewley2001} or the \citet{kauffmann2003} diagnostics ($\log({\rm{[OIII]_{\lambda5007}/H\beta}}$) = -0.75, $\log({\rm{[NII]_{\lambda6584}/H\alpha}}$) = $-0.39$).  In Section 2, we describe our X-ray and near-infrared ground-based observations and data analysis, followed by a description of our results in Section 3.  In Sections 4,5, and 6, we discuss various possible explanations for the origin of the X-ray, mid-infrared, and near-infrared emission in ~\gal\, including obscured star formation and a deeply embedded AGN.  In Section 7 we analyze the multi wavelength spectral energy distribution of ~\gal.  In Section 8, we discuss the implications of our results, and summarize our findings in Section 9.

We adopt a standard $\Lambda$CDM cosmology with $H_0=70$~km~s$^{-1}$~Mpc$^{-1}$, $\Omega_{\textrm{M}}=0.3$, and $\Omega_\Lambda=0.7$.

\section{Observations and Data Reduction}
\subsection{XMM Observations}
\label{XMMobservations}

\gal\ was observed for 10~ks twice by \textit{XMM-Newton}, 9 May 2013 (obsid=0721900401) and 11 May 2013 (obsid=0721900601), as part of a pilot follow-up study (Cycle 12 GO, prop. 072190) using \textit{XMM-Newton} to investigate the nature of the red {\it WISE} sources presented in \citet{satyapal2014}. We note that a total of 4 targets were observed in this program.  One of the targets (SDSS~J012218.11+010025.8) was a merging system for which we obtained follow-up {\it Chandra} observations.  The {\it XMM} and {\it Chandra} observations of this target will be published as a larger sample of X-ray identified dual AGNs \citep{satyapal2016}.The 4th target was observed with \textit{XMM-Newton} (SDSS~J082240.29+034546.5), but the {\it WISE} photometry from the ALLWISE release of the {\it WISE} catalog showed that the magnitudes from the previous release were erroneous, and the source actually does not have updated red colors that meet the AGN criteria adopted in Satyapal et al.~2014. The observations presented here are thus the second of only two isolated bulgeless galaxies from the sample from \citet{satyapal2014} for which follow-up X-ray observations were obtained.

\begin{figure}
\noindent{\includegraphics[width=7.8cm]{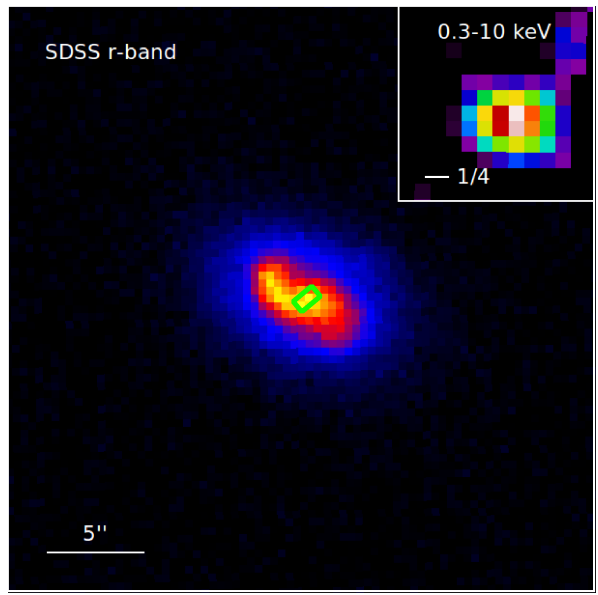}}
\caption{SDSS r-band image with insert of 0.3-10 keV X-ray image at 1/4 scale. X-ray image binned by 64 and smoothed with a 3-pixel Gaussian for clarity.  The LBT extraction aperture is overlaid in green.}
\label{xrayimage}
\end{figure}

We reduced the data with the \textit{XMM-Newton Science Analysis Software} (SAS), version 13.0.0, using the most up to date calibration files. We filtered our event files using \texttt{evselect} for "good" (\texttt{FLAG}=0). Examination of the pn data reveals a faint source associated with \gal\ (see Figure~\ref{xrayimage}).  To obtain better photon statistics, we merged the (cleaned) event files from all cameras and both observations into a single event file using the \texttt{merge} task.  We created a 0.3-10~keV image of \gal, binned by a factor of 64, and we used \texttt{eregionanalyse} to extract counts and compute the background-subtracted count rate for the X-ray source.  We used a circular extraction region centered on \gal\ with a radius of 200 pixels ($10\arcsec$), which we found to be the optimum extraction radius for maximizing the signal-to-noise.  The encircled energy factor for this extraction region is 0.62.  We used a 1200 pixel ($1\arcmin$) radius background region which we placed on a neighboring, source-free area on the image.

We used a weighted Galactic total hydrogen column density of $N_\mathrm{H}=1.20\times10^{20}$~cm$^{-2}$, derived from the \textit{Swift} Galactic $N_\textrm{H}$ tool,\footnote{\url{http://www.swift.ac.uk/analysis/nhtot/index.php}} which is based on the work of~\citet{Willingale+2013} that appends the molecular hydrogen column density $N_\textrm{HII}$ to the atomic hydrogen column density $N_\textrm{HI}$ from the Leiden-Argentine-Bonn (LAB) 21-cm survey~\citep{Kalberla+2005}.

\subsection{Large Binocular Telescope Observations}
%\textcolor{blue}{Barry, please check carefully.}.

In order to search for obscured star formation, or broad lines obscured in the optical, we obtained near-infrared ground based spectroscopy with the LUCI- 1 instrument \citep{seifert2003,seifert2010} mounted on the \textit{Large Binocular Telescope (LBT)} on March 28, 2015.
The LBT employs two 8.4 m mirrors on a single mount.
LUCI-1 is a near-IR imager/spectrograph for the LBT with wavelength coverage of $0.85-2.4$ $\mu$m ({\it zJHK} bands) in imaging, long-slit and multi-object spectroscopy modes.  The spectrum of  SDSS~J1224+5555 was taken with the N1.8 camera, centered on the coordinates R.A. = $12^{\rm h}24^{\rm m}34.^{\rm s}64$, decl. = $+55\deg 55\arcmin 22.619\farcs$ and a position angle of $40\deg$.  We employed a 1{\arcsec}.0 wide $\times$ 3{\arcmin}.9{\arcsec} long slit, and the G200 grating with the HKspec filter.  At the distance of \gal,  $1\farcs0$ corresponds to $\approx1~kpc$.  Using calibration Ne and Ar arc lines, we measure an average spectral resolution of 0.0017~$\micron$ per pixel, 
or ${\it R}$ $\sim$ 858-1376 over this wavelength range. The total integration time was 30 minute, split into six individual exposures.  As the target galaxy is significantly smaller than the slit length, observations were done using an {\tt AB} pattern of nodding along the slit.  Thus, every exposure contains
the science target. The A0V type star HIP56736 was observed before the target at similar airmass, to remove telluric features.  

\indent The LUCI-1 data were reduced using a set of custom {\tt IRAF} scripts.  Briefly, first the science and A0V telluric star exposures were trimmed and flatfielded; 
then, science and telluric star exposures closest in time were pair subtracted to remove background IR flux and nightsky emission lines.  Next,  wavelength calibration was done 
by fitting a 3rd order {\it spline3} polynomial to 37 nightsky lines identified in the science exposures.  The fitting yielded an RMS~=~$7.68\times10^{-5}$~$\micron$ over the 
entire wavelengh range.  In the detector X-Y pixel reference frame LUCI-1 spectra are slightly tilted along the X-axis and show non-negligible curvature in the Y-direction.  
To correct for both of these effects, the {\tt IRAF} tasks {\tt REID}, {\tt FITC}, and {\tt TRANSFORM} were used to straighten and rectify all science and A0V telluric star 
exposures via the wavelength calibration applied to the central 5 rows.  The pair subtracted science frames were then shifted in Y and combined into a single frame in order
to improve the signal-to-noise (S/N) for extracting a one-dimensional spectrum. The {\tt IRAF} task {\tt APALL} was used to extract a one-dimensional spectrum using 
a 2 pixel (or 0{\arcsec}.5 in angular size) diameter aperture from the combined science frame.  At the redshift of the target, 2 pixels corresponds to an aperture of
500 pc. The angular size of the aperture was chosen based on the measured full width at half maximum (FWHM) of stars in the acquisition images, the FWHM in the spatial direction
of the telluric star frames, and continual monitoring of the wavefront sensor of the guide camera.\\
\indent Once the one-dimensional, wavelength calibrated science and telluric spectra were extracted, corrections and flux calibration were applied using the {\tt IDL}-based
software {\tt XTELLCOR GENERAL}.  This package uses the spectrum of an A0V star observed close in time and airmass to the science target and a high-resolution spectrum
model of Vega to create a telluric correction spectrum free of stellar absorption lines in order to then remove telluric absorption features in the data \citep{vacca2003}.
{\tt XTELLCOR GENERAL} also flux calibrates the science data using the known $(B-V)$ color of the A0V star, and assuming there are not significant slit losses in the 
observations.  The final telluric corrected and flux calibrated data were then used for analysis in the paper.

\section{Results}

\subsection{X-ray Results }

Using the source and background extraction regions described in \S\ref{XMMobservations}, we find $25\pm8$ counts between 0.3-10~keV for \gal, a $3\sigma$ detection.  The effective exposure time for the merged event file was 20~ks, yielding a background-subtracted count rate of $(1.2\pm0.4)\times10^{-3}$~cnt~s$^{-1}$.  Because we do not have enough counts to reliably estimate a hardness ratio, we estimated hard (2-10~keV) X-ray fluxes assuming a typical AGN power-law X-ray spectrum with index $\Gamma=1.8$, and using WebPIMMS.\footnote{\url{https://heasarc.gsfc.nasa.gov/cgi-bin/Tools/w3pimms/w3pimms.pl}}  The observed hard X-ray luminosity (assuming no intrinsic absorption) is $L_\mathrm{2-10~keV}=(1.1\pm0.4)\times10^{40}$~erg~s$^{-1}$. If we assume a value for the intrinsic absorption of $N_\mathrm{H}=10^{21}$~cm$^{-2}$, we estimate an unabsorbed hard X-ray luminosity of $L_\mathrm{2-10~keV}=(1.6\pm0.5)\times10^{40}$~erg~s$^{-1}$.  If this source is nearly Compton-thick ($N_\mathrm{H}\approx10^{24}$~cm$^{-2}$; see \S\ref{obscuredagn}), the unabsorbed hard X-ray luminosity is  $L_\mathrm{2-10~keV}\sim3\times10^{42}$~erg~s$^{-1}$. We note that the X-ray data alone is not sufficient to constrain the column density.

\subsection{Near-infrared Spectrum }

\begin{figure}
\noindent{\includegraphics[width=8.7cm]{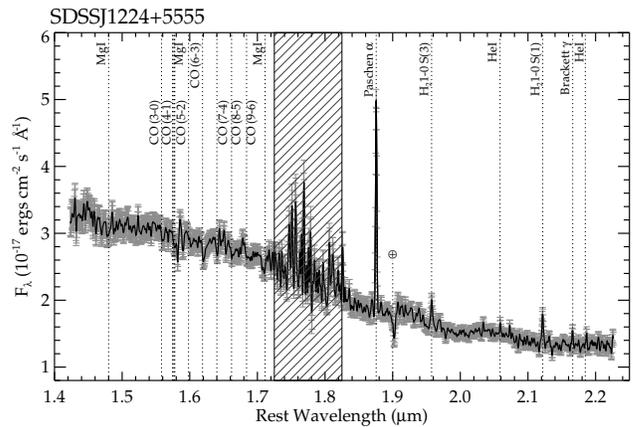}}
\caption{The LBT near-infrared spectrum of the nuclear source corrected for redshift, with labels for features detected. Also noted is a strong telluric absorption feature at ~$1.9~ \mu$m from imperfect telluric corrections, indicated by the encircled plus line.  The diagonally hatched lines note a region of strong absorption due to the Earth?s atmosphere.}
\label{irspectrum}
\end{figure}

\begin{figure}
\noindent{\includegraphics[width=8.7cm]{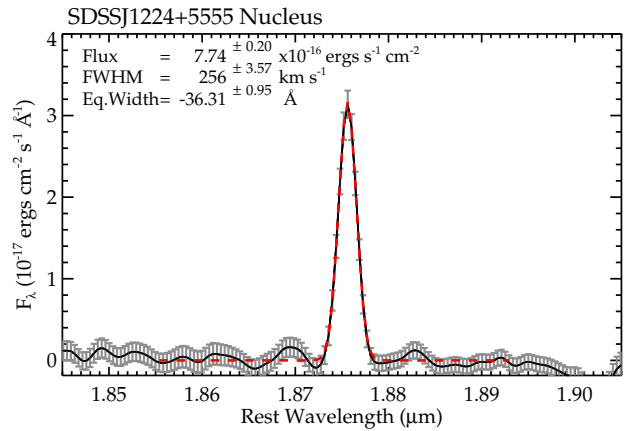}}
\caption{The near-infrared spectrum centered on the Pa$\alpha$ line .  The red line denotes the fit to the Pa$\alpha$ line, while the gray denotes the RMS noise.}
\label{paspectrum}
\end{figure}

In Figure~\ref{irspectrum}, we plot the 1D near-infrared spectrum corresponding to the position of the near-IR unresolved nuclear source. We detect a prominent Pa$\alpha$ line at $\lambda$=1.876{~\micron}. The near-infrared spectrum centered on the Pa$\alpha$ line  is shown in Figure~\ref{paspectrum}. The total Pa$\alpha$ flux within the 0.5$\arcsec$ aperture is $F_\mathrm{Pa\alpha}$=$(7.74\pm0.20)\times10^{-16}$~erg~cm$^{-2}$~s$^{-1}$ with a FWHM, corrected for the instrumental resolution, of $256\pm3.57$~km/s.  There is no evidence for any broad component to the emission line, and no high excitation lines were detected. We report approximately  $3\sigma$ detections of the $H_2~1-0S(3)~1.957{\micron}$ and $H_2~ 1-0S(3)~2.121{\micron}$ with fluxes of $(1.37\pm0.39)\times10^{-16}$~erg~cm$^{-2}$~s$^{-1}$ and $(1.26\pm0.29)\times10^{-16}$~erg~cm$^{-2}$~s$^{-1}$, respectively.  The  $3\sigma$  upper limit to the [FeII] 1.644~\micron\ flux is $F_\mathrm{[FeII]}\lesssim6.27\times10^{-17}$~erg~cm$^{-2}$~s$^{-1}$.
The  $3\sigma$  upper limit to the Br$\gamma$ flux is $F_\mathrm{Br\gamma}\lesssim9.32\times10^{-17}$~erg~cm$^{-2}$~s$^{-1}$ .

 \renewcommand{\baselinestretch}{1.00}

\section{The Origin of the X-ray Emission in ~\gal}
\label{origin}
\
The derived observed X-ray luminosity (uncorrected for intrinsic absorption) of \gal\ of $L_\mathrm{2-10~keV}=(1.1\pm0.4)\times10^{40}$~erg~s$^{-1}$ is low compared to powerful AGNs, and there are insufficient counts to fit the X-ray spectrum and constrain the intrinsic absorption and the nature of the nuclear X-ray source.  This leaves open the possibility that the X-ray emission results from the integrated emission from a population of X-ray binaries. We explored this possibility by estimating the contribution of X-ray binaries to the observed X-ray emission using the star formation rate (SFR) expected from a similar aperture. The global galaxy-wide SFR from the Max Planck Institut fur Astrophysik/Johns Hopkins University (MPA/JHU) collaboration\footnote{\url{www.mpa-garching.mpg.de/SDSS/}}, which follows~\citet{brinchmann2004} with photometric corrections from~\citet{salim2007} is 2.6~$M_\sun$~yr$^{-1}$.  However, since ~\gal\ is a late-type galaxy, it is possible that there is significant extinction at optical wavelengths and that the SFR based on the H$\alpha$ luminosity is significantly underestimated.  We explored this possibility using the observed Pa$\alpha$ line flux from the {\it LBT} spectrum.  The H$\alpha$  flux from the MPA catalog for \gal\ is  $F_\mathrm{H\alpha}$=$(7.735\pm0.0797)\times10^{-15}$~erg~cm$^{-2}$~s$^{-1}$. Since the SDSS H$\alpha$ flux corresponds to a 3~$\arcsec$ circular aperture, larger than the  $1.5\arcsec\times0.5\arcsec$ extraction aperture of the {\it LBT} spectrum, we can get a rough estimate of the extinction toward the recombination line emission by scaling the fluxes assuming a uniform surface brightness.  Scaling by the ratio of aperture areas, the aperture-corrected observed H$\alpha$ to Pa$\alpha$ line flux ratio for \gal\ is 0.70.  Assuming an intrinsic ratio of 7.82 for galaxies with 12+log(O/H)$>$8.35 \citep{osterbrock2006} and adopting an extinction curve with differential reddening of  k(H$\alpha$)-k(Pa$\alpha$)=2.08 f\citep{landini1984}, we obtain a color excess of E(B-V) of 1.255 \footnote{The color excess is calculated using the formula
$\frac{F_{int}(H\alpha )}{F_{int}(Pa\alpha)}=\frac{F_{int}(H\alpha )}{F_{int}(Pa\alpha)}10^{-0.4(k(H\alpha)-k(Pa\alpha))E(B-V)}$} .   Assuming a Milky Way-like extinction curve (R$_{V}$=3.1), we calculate only a moderate extinction of A$_{V}$ of 3.89 toward the ionized gas.  Since the $1\farcs0$ slit width corresponds to  $\approx1~kpc$ at the distance of \gal\, it is likely that the bulk of the ionized gas emission is contained within the {\it LBT} aperture.  The aperture correction used above therefore is likely to significantly underestimate the H$\alpha$ flux in the $1\farcs0$ slit.  The extinction is therefore likely to be lower than the value calculated here. 

If we use the upper limit to the  Br$\gamma$ flux, we can estimate a upper limit to the extinction.  Assuming an intrinsic Pa$\alpha$ to Br$\gamma$ flux ratio of 12.5 \citep{osterbrock2006}, the observed line flux ratio corresponds to an A$_{V} < 5.5$. Thus the recombination line fluxes detected from ~\gal\ together imply that the extinction toward the nuclear star forming regions is not excessively large. Indeed the SFR derived using the far-infrared $\mu$m luminosity, a widely used indicator of the unobscured SFR in both star forming galaxies and AGNs, adopting the value from the recent catalog by \citet{ellison2016}, is $\sim~2.5~M_\sun$~yr$^{-1}$, in excellent agreement with the MPA value, strongly suggesting that there is no significant embedded star formation in \gal\ .  Thus both the near-infrared spectra and the far-infrared IRAS photometry suggest only moderate extinction toward the star forming region in \gal. We note that \gal\ is not detected at 1.4 GHz by the NRAO VLA Sky Survey (NVSS\footnote{The National Radio Astronomy Observatory is a facility of the National Science Foundation operated under cooperative agreement by Associated Universities, Inc.})or FIRST (Faint Images of the Radio Sky at Twenty-Centimeters\footnote{http://sundog.stsci.edu/}) survey. The SFR implied by the upper limit  \citep{Bell2003} of the 1.4 GHz flux is $< 7.2~M_\sun$~yr$^{-1}$, fully consistent with MPA value , strongly suggesting that there is no significant deeply embedded star formation in \gal.

To calculate the predicted X-ray emission from XRBs, we used the relationship between stellar mass, star formation rate, and X-ray emission given in \citet{lehmer2010}.  Given $\log{M_\star/M_\sun}=10.3$ and SFR~=~2.5~$M_\sun$~yr$^{-1}$, the predicted 2-10~keV luminosity due to XRBs is $5.9\times10^{39}$~erg~s$^{-1}$, a factor of $\sim 3$ times lower than the apparent X-ray luminosity of ~\gal\ of $L_\mathrm{2-10~keV}=(1.6\pm0.5)\times10^{40}$~erg~s$^{-1}$.  However there is a considerable scatter on the \citet{lehmer2010} relation (0.34 dex), indicating that the observed X-ray luminosity is not significantly above the scatter of the relation.

\begin{figure}
\noindent{\includegraphics[width=8.0cm]{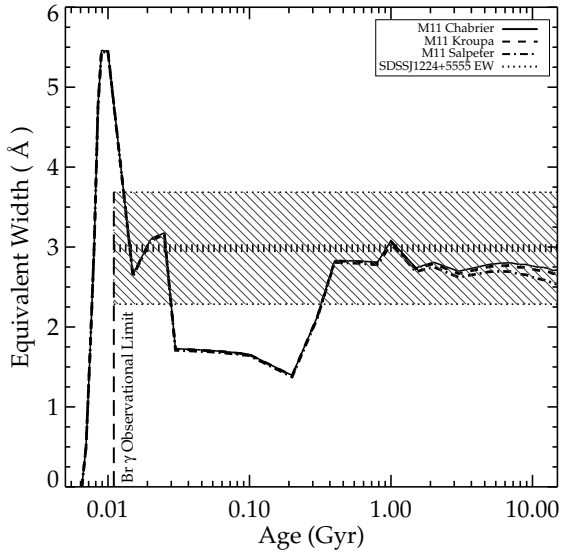}}
\caption{CO equivalent width vs. age. The horizontal dashed line is the measured EW of CO(6-3) and the the dotted grey lines above and below (and the diagonal hatched lines) indicate the EW error. The upper limit to the Br$\gamma$ EW is indicated by the dashed vertical line. These two constraints imply a stellar population age greater than $\sim$10 Myr.  }
\label{Fig:COEW}
\end{figure}

To further explore the possibility that the X-ray emission is of stellar origin, we investigated the age of the nuclear stellar population to determine if it was consistent with the existence of a large population of X-ray binaries. The relative numbers of LMXB and HMLXBs in a galaxy is strongly dependent on the specific star formation rate ($SFR/M_{star}$) \citep{grimm2002,gilfanov2004}. Using the SFR of $\sim~2.5~M_\sun$~yr$^{-1}$ and the galaxy stellar mass of $\log{M_\star /M_\sun}$=10.3, the specific SFR for ~\gal\ is  approximately 2 times higher than the Milky Way\citep{grimm2002}.  For this specific star formation rate, HMXBs are expected to outnumber LMXBs at the highest luminosities by a factor of 4 \citep{gilfanov2004}. If the X-ray luminosity of ~\gal\ is attributable solely to X-ray binaries, the underlying host galaxy should be dominated by a young stellar population, when the population of HMXBs is expected to be high.  Using the near-infrared spectrum of ~\gal, we investigated the extinction-insensitive age of the stellar population within the central 500~pc to see whether it is consistent with a population of HMXBs. The {\it H}-band is dominated
by the presence of stellar absorption lines (see \citet{Rayner09} for examples). 
The strongest feature within that wavelength range
is the CO (6-3) transition at 1.6189 $\micron$.  The CO bandhead is present in late-type giant stars, as
well as younger red supergiants.  The depth of the bandhead is known to vary with age and metallicity,
providing a way to constrain the ages of the stellar populations \citep[e.g.,][]{Origlia93,Oliva95,Origlia97}. We used the \citet{Maraston11} (hereafter M11) set of intermediate-high resolution stellar population models which include empirical stellar spectra.  At the near-IR wavelengths
of interest, the population models are sampled at 5{\AA} resolution, similar to the LUCI-1
data.  The M11 near-IR models use the Pickles empirical library \citep{Pickles98} which covers
a broad range of evolutionary phases from O-type stars and supergiants, to M dwarfs.
We convolved the M11 models to match the instrumental resolution of the LUCI-1 observations.
Then the CO (6-3) bandhead was measured in both the M11 models and LUCI-1 data using
the prescribed definition and methodology from  \citet{Origlia93}. Currently,  M11 employs
 solar metallicity and include Kroupa, Chabrier, and Salpeter IMFs.  In Figure~\ref{Fig:COEW}, we plot
the equivalent widths (EW) of the CO bandhead for the three different M11 instantaneous starburst models as a function of time.  The
horizontal line is the measured EW of SDSSJ1224+5555.  As can be seen, the evolution of the CO (6-3) bandhead is
age degenerate, since the Red Supergiant phase (t $<$ 20 Myr) and Red Giant phase (t $>$ 300 Myr)
show a marked increase in the EWs.  The measured EW of SDSSJ1224+5555 intersects several points in the
temporal evolution of the models. A commonly employed method to break this degeneracy is to use
recombination emission lines to constrain the ages. The equivalent width of the recombination lines are strongly dependent on age, showing a steep decline as the most massive stars evolve off the main sequence.  We applied the Starburst 99 (SB99) star-formation models to the
spectra of SDSSJ1224+5555 \citep{Leitherer95a,Leitherer14}.  In Figure~\ref{Fig:COEW}, we show the upper limit to the Br-$\gamma$ EW for ~\gal. The near-infrared spectrum therefore suggests that the underlying stellar population of the nuclear region in ~\gal is older than 10 Myr.  For solar metallicity galaxies, the peak in the number of  bright HMXBs ($L_X >10^{36}$~erg~s$^{-1}$  is approximately 5 Myr after the burst (see Figure 1 in \citet{linden2010}), decreasing to below 1 HMXBs for a starburst of $10^6$~$M_\sun$.  While the age implied by the near-infrared spectrum of ~\gal\ is not consistent with the peak in the HMXB population, we cannot rule out the possibility that the X-ray emission can be produced by a small number of extremely luminous HMXBs.

% \subsection{Estimating the Ages of the Stellar Population}

%

\section{The Origin of the Mid-Infrared Emission}
\label{obscuredagn}

\begin{figure}
\noindent{\includegraphics[width=8.7cm]{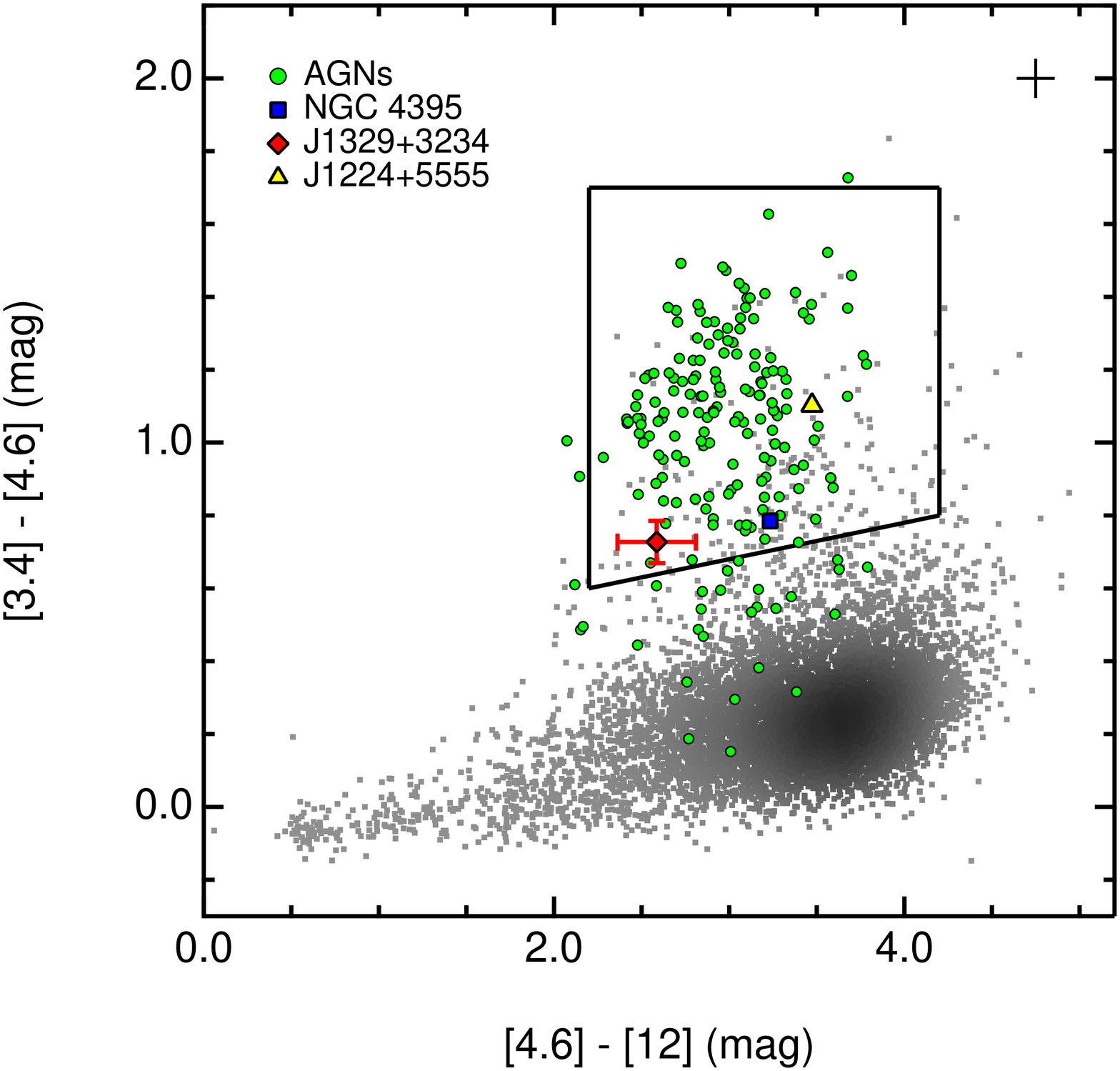}}
\caption{$W1$-$W2$ color versus the $W2$-$W3$ color for AGNs, optically star forming galaxies from SDSS (grey dots)}, and \gal, with J1329+3234 from \citet{secrest2015} and NGC 4395 included for comparison. The AGN region from \citet{jarrett2011} is also shown as the solid polygon region, and the typical uncertainty for the AGN colors is given by the black cross at the top right.
\label{colorcolorplot}
\end{figure}

While the X-ray observation alone does not require the presence of an AGN, the mid-infrared emission from ~\gal\ is consistent with an AGN that dominates the mid-infrared luminosity.  In Figure~\ref{colorcolorplot}, we plot the $W1$-$W2$ versus $W2$-$W3$ colors of ~\gal\ together with the 3-band AGN demarcation region from \citet{jarrett2011}.  We also plot the mid-infrared colors of NGC 4395, J1329+3234 from \citet{secrest2015}, the hard X-ray selected sample of 184 AGNs from \citet{secrest2015}, and galaxies classified as optically normal from the SDSS Data Release Eight.  Optical emission line ratios were taken from the MPA/JHU DR8 catalogs \footnote{\url{https://www.sdss3.org/dr10/spectro/galaxy_mpajhu.php/}} and optically non-active galaxies were identified using the demarcation between AGNs and star-forming galaxies from \citet{stasinska2006}. As can be seen, \gal\ falls squarely in the center of the AGN region, and is significantly redder than J1329+3234 and the unambiguous broad line AGN NGC 4395. 

For samples of \textit{known} AGNs, \textit{WISE} color selection has been shown to be extremely reliable for identifying AGNs.  For example, \citet{stern2012} show that for a sample of \textit{Spitzer}-identified AGNs from the \textit{Spitzer}-COSMOS field \citep{sanders07}, a simple one-color \textit{WISE} color cut of $W1$-$W2>0.8$ identifies AGNs with a reliability of 95\%, and the two-color \textit{WISE} selection criterion of \citet{mateos2015} recovers luminous ($L_\textrm{2-10~keV}>10^{44}$~erg~s$^{-1}$) hard X-ray-selected type-1 AGNs with a reliability of 97\%.  This same two-color selection criterion was also shown to by itself recover 98\% of AGNs in a magnitude-limited ($g<20$) sample from the SDSS-DR12Q catalog below a redshift of $z<2$ \citep{secrestb2015}.  Thus it is clear that for powerful AGNs, mid-infrared color selection is very effective in identifying AGNs.

We note that of all the optically classified SDSS galaxies plotted in Figure~\ref{colorcolorplot}, only 1.8\% have mid-infrared colors within the AGN demarcation region shown in Figure 5, demonstrating that mid-infrared colors of optically normal galaxies are rarely within the mid-IR AGN box. We also note that the absence of optical signatures of activity in these galaxies does not of course rule out the presence of an AGN.  We searched the 3XMM-DR5 catalog \citep{rosen2016}, which contains X-ray source detections drawn from 7781 {\it XMM-Newton} EPIC observations to see if any of the optically normal galaxies in the mid-IR AGN box in Figure ~\ref{colorcolorplot} have associated X-ray point sources.  To guard against any potential bias that might be introduced by pointed observations of the targets, we conservatively selected only sources that were greater than $1\arcmin$ from the {\it XMM-Newton} aim-point to ensure that any detected point sources would be serendipitous discoveries. There are 132 optically normal star forming galaxies in the mid-IR selection box plotted in Figure~\ref{colorcolorplot} that have X-ray sources in the 3XMM-DR5 catalog, 94\% of which have $\log{L_\textrm{2-10~keV}} > 42 $ [erg~s$^{-1}$, strongly suggesting that they host optically hidden AGNs.  Thus, the sources inside the mid-infrared color selection region are overwhelmingly AGN, and the mid-infrared colors of \gal are therefore highly suggestive of an AGN.

While the red mid-infrared color of \gal\ is highly suggestive of AGN activity, it is of course possible that red mid-infrared colors arise from dust heated by stellar processes. There have been a handful of low metallicity blue compact dwarfs (BCDs) with extreme red mid-infrared colors, raising the possibility that there is a similar origin for the hot dust in \gal. For example, \citet{izotov2011} find, from a sample of $\sim5000$ SDSS galaxies with \textit{WISE} colors, 4 dwarf galaxies with extreme \textit{WISE} colors ($W1$-$W2>2$~mag).  Since the hardness of the stellar radiation increases with decreasing metallicity \citep[e.g.][]{campbell1986}, and BCDs contain significant star formation, the dust in BCDs can potentially be heated to higher temperatures than is typically seen in starburst galaxies. However, \gal\ has an above solar metallicity, suggesting that metallicity does not play a role in explaining the mid-infrared colors. Moreover, the four galaxies from the \citet{izotov2011} sample occupy a separate region in mid-IR color space than \gal, with much redder colors ($W1$-$W2$=2.13-2.37; $W2$-$W3$=3.58-4.76), indicating considerably more emission at longer wavelengths from dust heating due to starbursts. We also point out that the analysis of the equivalent width of the CO bandhead and the absence of the Br$\gamma$ line discussed in section \S\ref{origin} suggests that the underlying nuclear stellar population in \gal\ is older than 10~Myr, and therefore not consistent with the hottest dust that can be produced by star formation alone. 

\par
Figure ~\ref{colorcolorplot} is suggestive that the mid-infrared emission is dominated by an AGN which is obscured in the optical. An excess in the mid-infrared emission expected from star formation is also suggested by the far-infrared emission from ~\gal. If we assume that the galaxy's far-infrared emission is produced by stars alone and that even if the 
galaxy does host an AGN, its contribution to this wavelength range is negligible \citep{deGrijp1985,deGrijp1987,Alton1998,Coziol1998,Netzer2007,Mullaney2011,Rosario2012},
 we predicted the W2 luminosity expected from star formation assuming that it has a similar spectral energy distribution (SED) to the prototypical starburst, M82.  The W2 to 60 $\mu$m flux ratio for M82 is $f_{\lambda,W2}$/$f_{\lambda,I60}$=0.54 \citep{polletta2007}, four times lower than the flux ratio in \gal\ ($f_{\lambda,W2}$/$f_{\lambda,I60}$=$2.24 \pm 0.20$), strongly suggesting that most of the emission in the W2 band is unlikely to be due to star formation and is most likely due to an obscured AGN. We note that the recent work by \citet{ellison2016} using the Herschel data shows that the IRAS-derived far-infrared luminosities are over-estimated in galaxies when a companion galaxy is nearby, as is the case for \gal\ . Using the Herschel data as a training set, they employ artificial neural networks (ANN) to generate a large public catalog of the far-infrared luminosities of a large sample of galaxies. The ANN value for the far-infrared luminosity from the \citet{ellison2016} catalog is a factor of ~2 lower than that predicted by IRAS, suggesting that the 60 micron flux from IRAS could be over-estimated by a factor of 2, possibly suggesting an even larger discrepancy in the  W2 to 60 $\mu$m flux ratio of a factor of 8 compared with M82.

\begin{figure}
\noindent{\includegraphics[width=8.7cm]{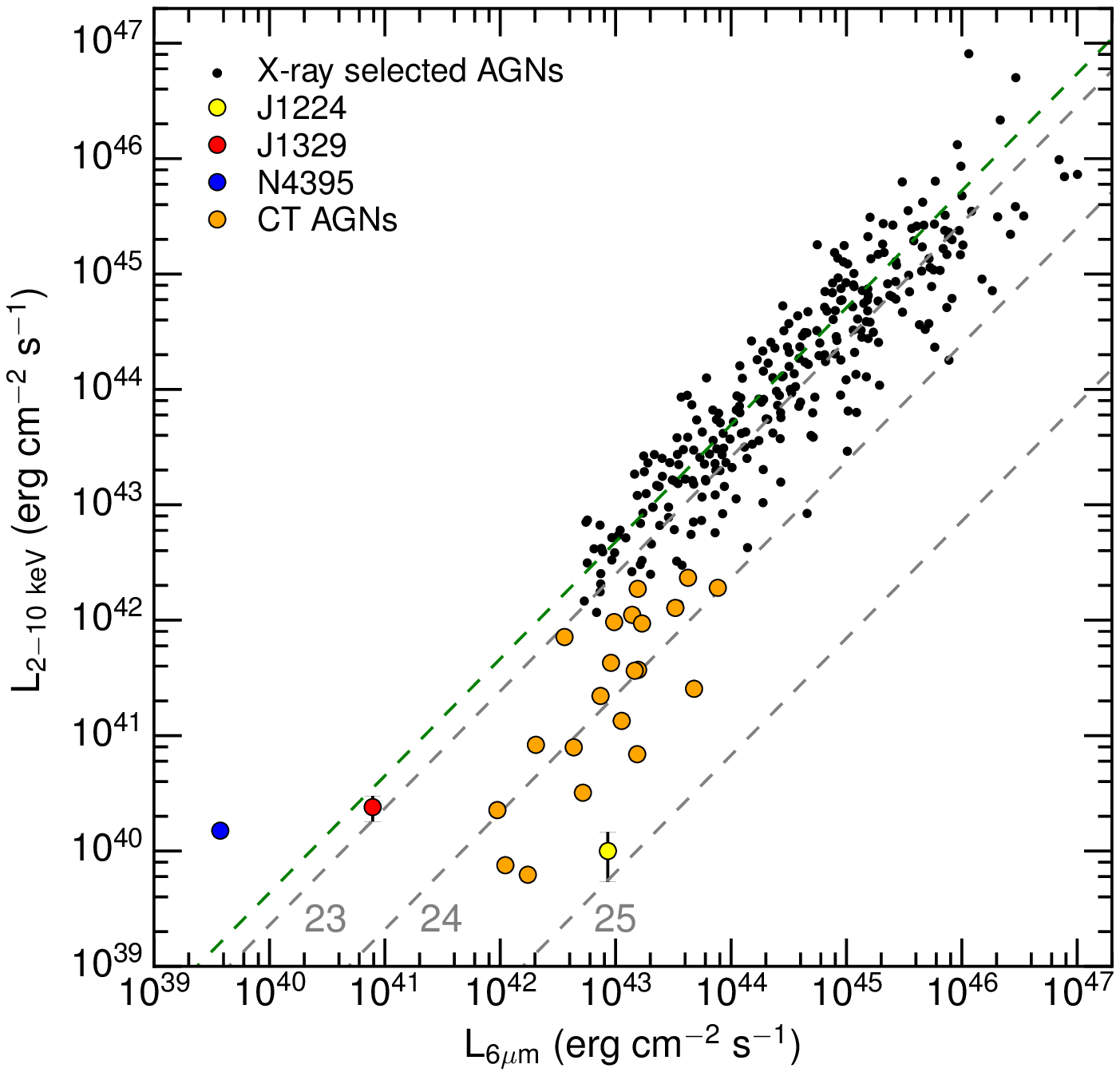}}
\caption{The observed 2-10~keV luminosity versus the 6~\micron\  luminosity for the hard X-ray selected sample of AGNs from \citet{secrest2015}, \gal, and the Compton-thick sources for which {\it XMM-Newton} observations are available from \citet{ricci2015}.  For comparison, we also plot NGC~4395, the archetypical bulgeless dwarf Seyfert 1 galaxy, as well as  J1329+3234, a bulgeless dwarf galaxy from \citet{secrest2015} that likely contains a massive black hole.  The dashed green line shows the linear regression between $L_\mathrm{6\micron}$ and intrinsic $L_\mathrm{2-10~kev}$ from \citet{mateos2015}. The distance below this line is used to plot the median predicted relation assuming an intrinsic absorption of $N_\mathrm{H}=10^{22}$, $10^{23}$, $10^{24}$, $10^{25}$~cm$^{-2}$ using a MYTorus model assuming $\Gamma=1.8$ and an inclination angle of $\theta=90\deg$ with X-ray scattering fractions that ranged from 0.01\%-1\%(gray dashed lines, where the exponent in the intrinsic $N_\mathrm{H}$ indicated at the base of each line).}
\label{lxlw2plot}
\end{figure}

Although the mid-infrared color of \gal\ is typical of dominant AGNs, its X-ray luminosity derived assuming only Galactic absorption is low compared to the hard X-ray selected sample of AGNs from \citet{secrest2015}, the majority of which have $N_\mathrm{H}\ <  10^{22}$~cm$^{-2}$, which ranges from  $\log{L_\textrm{2-10~keV}}=42.1-46.7$~[erg~s$^{-1}$]. The mid-infrared luminosity, thought to be re-emitted by the obscuring torus, and the AGN intrinsic 2-10 keV are known to follow a tight correlation over several orders of magnitude \citep[]{lutz2004,gandhi2009,mateos2015}. In Figure~\ref{lxlw2plot}, we plot the 6~\micron\ band luminosity verses the observed hard X-ray luminosity for the AGN sample described in \citet{secrest2015}, \gal, and the Compton-thick sources for which {\it XMM-Newton} observations are available from the 70 month {\it Swift/BAT} catalog from \citet{ricci2015}. We also plot the effect  of absorption on the local AGN relation using the MYTorus \citep{murphy2009} model assuming $\Gamma=1.8$ and an inclination angle of $\theta=90\deg$ with X-ray scattering fractions that ranged from 0.01\%-1\%(dotted gray lines). As can be seen, \gal\ is significantly under-luminous in the X-rays given its  6~\micron\ luminosity compared with the AGN sample from \citet{secrest2015} and, like the sources from \citet{ricci2015}, is located near the Compton thick regime . If we assume that the X-ray emission  and  6~\micron\ emission from ~\gal\ are due to an AGN, this suggests that the AGN is highly obscured, even in the X-rays.  Using the linear regression fit between $L_\textrm{2-10~keV}$ and $L_{6~\micron}$ for the AGN from \citet{mateos2015} (green dashed line in Figure~\ref{lxlw2plot}, and assuming an AGN power law slope of $\Gamma=1.8$, we estimate that the X-ray emitter in \gal\ is bordering on being Compton-thick, with $N_\mathrm{H}\approx3\times10^{24}$~cm$^{-2}$. For $N_\mathrm{H}\approx10^{24}$~cm$^{-2}$, the unabsorbed hard X-ray luminosity is $L_\mathrm{2-10~keV}\sim3\times10^{42}$~erg~s$^{-1}$.  We note that the value of $N_\mathrm{H}$ adopted must be viewed with some caution since it depends on the X-ray model assumed and does not include an extinction correction to the  6~\micron\ band luminosity, which may be significant for some Compton-thick sources  \citep[e.g.,][]{goulding2012}. Furthermore, we are assuming that the suppression of X-ray emission relative to the mid-infrared emission in \gal\ is due entirely by absorption and not intrinsic X-ray weakness, and that the contribution from star formation in the two bands is negligible. However, if there is an AGN in \gal\, it is likely that it is Compton-thick with $N_\mathrm{H}>10^{24}$~cm$^{-2}$. We adopt a value of .$N_\mathrm{H}\approx2\times10^{24}$~cm$^{-2}$ in this work.

\par

\section{The Origin of the Near-Infrared Emission}

The near-infrared spectral region offers access to several collisionally-excited forbidden transitions in highly ionized species, a robust indicator of an AGN, and since the extinction in the K band is roughly a factor of ten less than that in the optical, near-infrared spectroscopy potentially offers a less obscured view of the nuclear region. In addition near-infrared broad hydrogen recombination lines can also reveal an optically hidden AGN.  The {\it LBT} spectrum covers the wavelength of the [Si~VI]~1.963~$\micron$ coronal line, however no line was detected in the observed spectrum of \gal{}.   The absence of this line however is not unusual. Indeed, this line is frequently not detected even in optically confirmed Type 2 AGNs \citep[e.g.,][]{mason2015, riffel2006}. Even among a subsample of the {\it Swift}/BAT AGNs from the 70 month survey \citep{Baumgartner2013}, which represent the most powerful hard X-ray selected AGNs in the local Universe, only $\approx$ 40\% have detections in the [Si~VI] line in recent follow-up observations \citep{koss2016}.   If the extinction toward the AGN in ~\gal{} is very high, as the combined X-ray and mid-IR emission suggest, the absence of the [Si~VI]~1.963~$\micron$ line would therefore be expected.

%Similary, if \gal\ does harbor an obscured AGN, then the question arises why there is no evidence for a broad Pa$\alpha$ and/or Br$\gamma$ lines in the galaxy's {\it LBT} spectrum. We can predict the expected broad line flux based on the intrinsic X-ray luminosity  we estimated in \S\ref{obscuredagn} .   Using the relation between the intrinsic hard X-ray luminosity and the intrinsic broad  H$\alpha$ luminosity of type~1 AGNs from \citet{Terashima2000}, the predicted intrinsic  H$\alpha$ luminosity from the broad line region is $\sim5\times10^{42}$~erg~s$^{-1}$, which corresponds to an intrinsic Br$\gamma$ luminosity of ~$5\times10^{40}$~erg~s$^{-1}$.

%If the AGN is obscured by an intrinsic absorption  $N_\mathrm{H} = 2.2\times10^{24}$~cm$^{-2}$ (see \S\ref{obscuredagn}), as implied by the ratio of the X-ray and mid-IR luminosities, the visual extinction implied by this value of $N_\mathrm{H}$ is A$_{V} >1000$, assuming N$_{H}$=1.8$\times$10$^{21}$A$_{V}$ cm$^{-2}$ mag$^{-1}$ and R$_{V}$=3.1 \citep{Predehl1995}.Using a canonical extinction law \citep{landini1984}, the extinction at the wavelength of the ${Br\gamma}$ line is still A$_{Br\gamma}$$>~100$.  Therefore the lack of broad lines in the near-infrared spectrum is fully consistent with the scenario in which the AGN is obscured by the column density implied by the X-ray and mid-infrared results.  So the absence of broad lines in the infrared is still consistent with this 
%galaxy being a heavily obscured AGN, and the Pa $\alpha$ line that we do observe must come from the less obscured stellar region. 

 Similarly, there is no evidence for a broad Pa$\alpha$ or Br$\gamma$ line in \gal. However, the absence of broad recombination lines in the near-infrared is actually common even in optically identified type 2 AGNs. In a recent near-infrared spectroscopic study of local {\it Swift}/BAT AGNs
with $42 < log(L_\mathrm{2-10~keV}) < 45$ erg/s, \citet{onori2016} find that only 14 out of 41
Type 2 AGNs show broad Pa$\beta$ emission. If the extinction toward the AGN in ~\gal\ is very high, the absence of the broad recombination lines in the near-infrared spectrum would therefore be expected.

\par

Although the near-infrared spectrum of \gal\  does not show robust signatures of an AGN, we detected the $H_2~1-0S(3)~1.957~\micron$ and $H_2~ 1-0S(3)~2.121~\micron$ lines in the {\it LBT} spectrum.  These lines are detected frequently in both star forming and active galaxies with varying degrees of nuclear activity \citep[e.g.,][]{ mouri1994, rod2004, rod2005, riffel2006, riffel2013, mason2015}.  While there are multiple excitation mechanisms that can produce the emission lines, including UV fluorescence \citep{black1987}, shocks\citep{hollenbach1989}, and X-ray illumination \citep{maloney1996}, each of which will produce a different $H_2$ emission line spectrum, 
there is considerable ambiguity in determining the dominant excitation mechanism responsible for the line emission. Moreover, the same excitation mechanisms may be at play in both star forming galaxies and AGNs.  However, empirically there appears to be a separation between AGNs and star forming galaxies in the ratio of the near-infrared $H_2$ lines to the hydrogen recombination lines. In particular, several authors have noted a correlation between the [FeII]~1.257~$\micron$/Pa$\beta$ and the H$_2$~1-0~S(1)/Br$\gamma$ ratio in galaxies, with the lowest ratios seen in star forming galaxies \citep{larkin1998, rod2005, riffel2013}.  Using the largest sample of active and star forming galaxies from \citet{riffel2013}, the line ratios are found to be separated by nuclear activity class according to the following demarcations: [Fe II]/ Pa$\beta$ $<$ ~0.6 and $H_2$ /Br$\gamma$ $<$ 0.4 for star forming galaxies; 0.6 $<$ [Fe II]/ Pa$\beta$ $<$ 2  and 0.4 $<$ $H_2$ /Br$\gamma$ $<$  6 for AGNs;  and for LINERs, [Fe II]/ Pa$\beta$ $>$ 2 and $H_2$ / Br$\gamma$ $>$ 6.  Using the flux of the $H_2~1-0~S(1)$ line and the upper limit to the Br$\gamma$ flux, the $H_2~1-0~S(1)$/Br$\gamma$  flux ratio of ~\gal\  is $>$ 1.35, greater than the region generally occupied by star forming galaxies. This raises the possibility the extended emission probed by the [FeII] and $H_2$ emission is influenced by the AGN, even though the broad line is obscured in the near-infrared. However we point out that there is some overlap in these line ratios between the AGN and star forming regions in the sample studied by  \citet{riffel2013}, indicating that the interpretation of the emission lines ratios from such low ionization species is ambiguous (see also \citet{smith2014}).

\section{Multiwavelength SED}

%Jess, I don't know if this table makes sense to include. Did you do an iterative chi^2 to arrive at your best fit?  Do the galaxy templates also include a variable extinction, and if so, can you exclude them if you add that in?  Adding the AGN template would naturally improve chi^2 since you are increasing the number of parameters in the fit.  I am not sure how degenerate these models are and I think we need to be careful on what we say here
%

\begin{figure}
\noindent{\includegraphics[width=8.0cm]{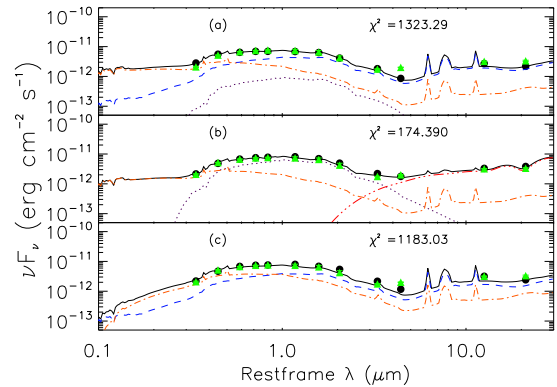}}
\caption{The observed photometry of SDSS J1224+5555 (green triangles) compared to model photometry (black circles, solid black line) created using combinations of the E (purple dotted line), Sbc (blue dashed line), Im (orange dot-dashed line) and AGN (red triple dot-dashed line) templates from \citet{assef2010}. Panel (a): Best fit model of a linear combination of the E, Sbc and Im templates assuming no additional extinction toward's the galaxy templates. Panel  (b): Best fit model using both the galaxy and AGN templates with variable extinction added to all the templates. Panel (c): Best fit model using only the galaxy templates but with additional extinction added as a free parameter. The $\chi^2$ value for the fit is shown in the upper right of each plot.}
\label{Fig:allpanels}
\end{figure}

The multiwavength UV to mid-IR spectral energy distribution (SED) of \gal\ supports the scenario of a highly obscured AGN, based on the empirically derived templates of \citet{assef2010}. These templates
consist of a set of three galaxy templates and one AGN template. The galaxy templates, E, Sbc and Im are based on templates 
from \citet{Coleman1980} with wavelength ranges extended using stellar models from \citet{Bruzual2003} and added dust and PAH 
components from \citet{Devriendt1999}. The AGN template is based on the average Type 1 AGN template from \cite{Richards2006} with 
efforts to remove host galaxy contamination and create a template for an unobscured AGN. This method has since been used to
successfully uncover highly obscured and even Compton-Thick AGN in several other works thus far 
\citep{Lansbury2014,Chung2014,Stern2014,hainline2014,chen2015,assef2015,Lansbury2015}. Figure \ref{Fig:allpanels} shows the 
best fit linear combination of the three galaxy templates to the observed SDSS, 2MASS and WISE photometry of the target galaxy, where the best fit is obtained through a $\chi^2$\ minimization.  The multi wavelength fluxes were obtained from the 
NASA/IPAC Extragalactic Database(NED)\footnote{https://ned.ipac.caltech.edu/}.
The maximum wavelength of the \citet{assef2010} templates is 30~$\mu$m; we therefore cannot use the IRAS 60 $\mu$m flux in our fits. As discussed in Section 5, the 60 $\mu$m flux for \gal\ is likely over-estimated and its use is in any case questionable.
While the SDSS $g-z$, 2MASS J-K$_{s}$, and WISE W1 bands are well fit by a combination of the Sbc
 and Im templates, the combined model 
flux is much lower than the observed fluxes in the W2-W4 bands. In fact, Table 2 of \citet{assef2010} shows that at z=0 and z=0.1, no galaxy 
template has a W1-W2 color greater than  0.2 (Vega). However when we add in the AGN template with a variable level of obscuration (red line) 
in Figure \ref{Fig:allpanels}, we see that the observed multiwavelength photometry of the galaxy is better fit with a combination of a young and old stellar population and highly obscured AGN (red line) dominating the WISE bands, with a $\chi^2$\ value almost a factor of 7 times lower than the best fit obtained using the galaxy templates alone. As an additional test, we investigated the effect of adding variable additional extinction to the galaxy templates alone to determine if a comparable fit can be obtained without invoking the presence of an obscured AGN contribution. However, as can be seen from Figure \ref{Fig:allpanels}, the best fit model cannot adequately fit the SED in the {\it WISE} bands, and results in a $\chi^2$\  value that is significantly worse than the model which includes an AGN (Figure \ref{Fig:allpanels}). We note that the improvement in fit when the AGN template is added cannot be attributed simply to having additional model parameters in the fit.  In the first model without an AGN, we achieve a $\chi^2$ of 1323.29, with 3 model parameters.  While the addition of an AGN template more than doubles the number of model parameters to 8, $\chi^2$ is reduced by a factor of $\sim8$, indicating that a model that includes an AGN template produces a significant improvement to the fit of the data.

We note that SED fitting of the multi wavelength photometry of galaxies can be highly degenerate. Furthermore, the \citet{assef2010} templates are based on the median SEDs of large samples of galaxies.  They therefore may not be appropriate to use on rare objects such as \gal.  Thus, while the above fits are suggestive of the presence of an obscured AGN, we do not consider them as providing robust proof of one.

%\begin{deluxetable*}{cccccc}
%\tablecolumns{6}
%\tablecaption{Normalization Constants for Assef Templates}
%\tablehead{\colhead{Fit Type} & \colhead{E} & \colhead{Sbc} & \colhead{Im} & \colhead{AGN} & \colhead{A$_{V}$}}
%\startdata
%Galaxy Only & 0 & 1.12$\times$10$^{-14}$ & 8.42$\times$10$^{-16}$ & 0 & 0 \\
%Galaxy+AGN & 4.85$\times$10$^{-15}$ & 0 & 1.10$\times$10$^{-15}$ & 4.78102$\times$10$^{-16}$ & 28.84 \\
%\enddata
%\label{Tab:constants}
%\end{deluxetable*}

\section{Discussion}

The analysis of \gal\ conducted in this work reveals: 1) an unresolved nuclear X-ray source consistent with but not exclusive to an AGN, 2) a near infrared spectrum that suggests that the age of the underlying nuclear stellar population is beyond the peak of the HMXB population, suggesting that it is unlikely that the X-ray emission is produced by HMXBs, 3) a nuclear stellar population age that is not consistent with heating the dust to the highest temperatures possible from star formation alone, 4) possible evidence that the extended  gas traced by the [FeII] and $H_2$ near-infrared emission lines is influenced by an AGN, 5) no evidence for significant deeply embedded star formation based on an analysis of the near-infrared and optical recombination line ratios and the far infrared and millimeter observations, 6) mid-infrared colors consistent with an AGN that dominates the mid-infrared luminosity, 7) an excess in the mid-infrared luminosity relative to the far-infrared luminosity, suggesting excess emission at mid-infrared wavelengths compared with that expected from star formation alone, and 8) a multi wavelength SED that cannot be well-fit by galaxy templates alone without the inclusion of an obscured AGN. While no one observation provides robust evidence for an AGN, the collective set of observations and analysis carried out in this paper suggests that \gal\ harbors a highly absorbed AGN.  

\subsection{Properties of the putative AGN in ~\gal}

If we assume that a highly absorbed AGN resides in ~\gal, we can estimate the black hole mass values for different accretion rate scenarios.   If we assume that the source is nearly Compton-thick ($N_\mathrm{H}\approx10^{24}$~cm$^{-2}$, the intrinsic hard X-ray luminosity is $L_\mathrm{2-10~keV}\sim3\times10^{42}$~erg~s$^{-1}$. Work from ~\citet{vasudevan2009} has shown that the bolometric correction factor, $\kappa_{2-10~keV}$, is a function of the Eddington rate. If we assume that \gal\ is a highly accreting AGN ($\lambda_{Edd} \sim 1$), the predicted bolometric correction factor is $\kappa_{2-10~keV}$=100, which yields a bolometric luminosity of the AGN of $L_\textrm{bol.}\sim3\times10^{44}$~erg~s$^{-1}$, and an estimated black hole mass of $M_\textrm{BH}\simeq2\times10^6~M_\sun$. If instead we assume that \gal\ is in the low accreting regime ($\lambda_{Edd} \sim 0.02$), the predicted bolometric correction factor is $\kappa_{2-10~keV}$=20, which yields a bolometric luminosity of the AGN of $L_\textrm{bol.}\sim6\times10^{43}$~erg~s$^{-1}$, and an estimated black hole mass of $M_\textrm{BH}\simeq2\times10^7~M_\sun$. In either scenario, \gal\ harbors a black hole of significant mass, well above the bulk of the distribution of plausible seed SMBH masses.

\subsection{Obscured AGNs in Bulgeless Galaxies?}
%{\textcolor{cyan}{Ryan, please look over carefully and add anything you see fit; your input here would be very valuable}}

The discovery of a growing number of optically quiescent galaxies in the low luminosity and low bulge mass regime with possible signatures of AGNs at X-ray and infrared wavelengths suggests that there may be an unexplored and significant population of obscured AGNs in the low luminosity regime, and that the impact on discriminating between seed formation scenarios may be significant. It is well known that the shape and the intensity of the cosmic X-ray background (CXB) cannot be reproduced successfully unless AGNs with intrinsic absorption of $N_\mathrm{H} > 10^{24}$~cm$^{-2}$ are a non-negligible fraction of the local AGN population \citep[e.g.,][]{comastri1995,gilli2007,treister2009}.  However, this population has been elusive in observational surveys, since identifying extremely obscured sources is challenging even in the local Universe.  Indeed, despite decades of multiwavelength observations from the X-ray to far-infrared wavelength region of the nearby ULIRG Arp 220, it is only through recent ALMA observations that an AGN obscured by a column density of  $N_\mathrm{H} = (0.6-1.8)\times10^{25}$~cm$^{-2}$ \citep{wilson2014,scoville2015} has been confirmed.  
\par
In galaxies that lack signatures of an AGN in their optical spectrum, X-ray observations have been very successful in identifying obscured AGNs.  However, at large column densities ($N_\mathrm{H} > 10^{23}$~cm$^{-2}$), the observed  2-10 keV luminosity is significantly affected by absorption.  Since the AGN intrinsic $L_\textrm{2-10~keV}$ and the mid-infrared luminosity of the obscuring torus are known to follow a tight correlation over several orders of magnitude \citep{lutz2004, gandhi2009,mateos2015,stern2015}, and the mid-infrared luminosity is less sensitive to obscuration, the selection of AGN with very low X-ray to mid-infrared  luminosity ratios has been used extensively to identify highly absorbed sources. Unlike J1329+3234, which appears to be only a moderately obscured AGN in a bulgeless dwarf galaxy, \gal\  shows signatures of heavier absorption.  While we have only obtained two follow-up {\it XMM-Newton} observations of bulgeless galaxies from the sample from \citet{satyapal2014}, it is interesting to note that \gal, which has a W2 luminosity almost two orders of magnitude greater than that of J1329+3234, shows evidence for heavier absorption.  Many studies suggest that the fraction of obscured AGN relative to the whole population decreases at the lowest and highest luminosities \citep[e.g.,][]{burlon2011,ballantyne2006,treister2006,lafranca2005,dellaceca2008,winter2009,brusa2010} .  At high luminosities, this is often attributed to a receding torus model \citep{lawrence1991}, in which the inner radius of the torus is set by the dust sublimation radius, which increases with AGN luminosity, a direct consequence of which would be an anti-correlation between the obscured AGN fraction and luminosity at high luminosities.  On the other hand, at low luminosities,  the torus obscuration may disappear, possibly because the clouds are generated by a disk-wind outflow, rather than accreted from the galaxy \citep{krolik1988}, resulting in less absorption of the X-rays. If we can increase the sample of bulgeless galaxies with both mid-infrared and X-ray observations, we can gain a better understanding of the column density distribution function and its dependence on luminosity and host galaxy properties in AGNs the local Universe.

\subsection{Challenges in Confirming AGNs in the Low Luminosity Regime - A Cautionary Tale}

The combined analysis of the X-ray, mid-infrared, and near-infrared emission, and the multi wavelength SED of ~\gal\ may suggest the presence of an obscured AGN, but definitive proof remains elusive.  The detection of AGNs in this low luminosity regime is challenging both because star formation in the host galaxy can dominate the optical spectrum and gas and dust can obscure the central engine at both optical and X-ray wavelengths.  For example, the detection of a compact radio source \citep{reines2012} coincident with a moderately absorbed  and variable nuclear X-ray point source \citep{reines2011,whalen2015}  in He 2-10 is highly suggestive of an AGN, despite its lack of AGN signatures in the optical. But the absorption corrected 2-10 keV X-ray luminosity observed by Chandra is a factor of ~$\approx$ 4 lower than the observed 2-10 keV luminosity of ~\gal\ and the  \textit{WISE} colors of He 2-10 (W1-W2=0.423; W2-W3=4.93) are typical of starburst galaxies (see Figure 26 in \citet{jarrett2011}), suggesting that the AGN does not dominate in the infrared.  Interestingly, He 2-10 shows no evidence for the high ionization [NeV] line in archival observations obtained with the high resolution spectrograph on board   \textit{Spitzer} and the ratio of the [NeIII]/[NeII] line flux ratio, which is sensitive to the hardness of the radiation field, is a factor of $\sim$ 10 lower than the average value in nearby starbursts \citep{engelbracht2008} suggesting that the putative AGN contributes minimally to the total bolometric luminosity of the galaxy.  Thus while mid-infrared color selection and X-ray observations at energies $<$ 10 keV are often powerful tools in uncovering optically obscured AGNs at higher luminosities, this is not the case in the low luminosity regime.  Indeed, none of the late-type and bulgleless galaxies that show evidence for the high excitation [NeV] line, a robust indicator of AGN activity, \citep{satyapal2007,satyapal2008,satyapal2009}, would be selected as AGNs based on optical spectroscopy or mid-infrared color selection by {\it WISE}.  

Similarly, as demonstrated in this work, the X-ray luminosities of weak AGNs will be low and can be comparable to, and therefore indistinguishable from, X-ray binaries in the host galaxy.  We should note that while several works have attempted to correct for the X-ray emission expected from X-ray binaries, there is considerable scatter in the relation, and there are numerous relations that have been derived using different samples of galaxies \citep[e.g.,][]{ranalli2003, Grimm2003,gilfanov2004b,lehmer2010,mineo2012} that show that the stellar mass, SFR, stellar population age, and mean metallicity can significantly affect the X-ray output of a given XRB population \citep{fragos2013}. Furthermore, the estimate of the nuclear SFR can vary significantly in different works, ranging from the use of optical or UV continuum measurements, to optical recombination line measurements, IRAS fluxes, to SED fitting, introducing additional uncertainties.  Thus caution must be exercised in determining a true excess in the X-ray emission expected over stellar sources in the low luminosity regime. Indeed, recent high spatial resolution integral field observations of the bulgeless galaxy NGC 3621 reveal that the narrow line emitting gas is spatially offset from the X-ray source previously identified as an AGN \citep{gliozzi2009} raising the possibility that the X-ray source is not associated with the AGN  \citep{Menezes2016}.  

These observations highlight the ambiguity associated with optical, X-ray, and mid-infrared color selection in the low luminosity regime, where both star formation and obscuration can both be at play and responsible for hiding the AGN. Indeed recent {\it NuSTAR} observations were required to provide definite proof of a low luminosity AGN in the Luminous Infrared Galaxy NGC 6286,  demonstrating that both heavy obscuration and a weak AGN can hide the AGN signatures at optical, infrared, and lower energy X-ray observations \citep{ricci2016}. There are a growing number of such low luminosity and optically unidentified AGNs that show evidence for heavy absorption based on {\it NuSTAR} observations \citep{Annuar2016}.  For very high absorption, it is even possible that mid-infrared selection can fail even in luminous AGNs as has recently been found in several Swift/BAT AGNs \citep{koss2016b}. Adding to the complexities described above, the time lag between the onset of accretion and its visibility at X-ray wavelengths and in the much more extended narrow line region gas, can also result in an incomplete census of AGN activity.  The impact of this effect is discussed recently by \cite{schawinski2015}  and may provide another explanation for the lack of optical signatures in some of the optically identified AGNs in the low bulge mass regime.  We note that while radio observations are insensitive to obscuration,  the radio emission in AGNs can be dominated by and indistinguishable from compact nuclear starbursts \citep{condon1991,delmoro2013}. Furthermore, only approximately 10\% of AGNs are radio loud and the radio loud fraction for low mass AGNs in particular is poorly known \citep{miller1990,stern2000}.Even broad emission lines in the optical spectrum, a robust indicator of an AGN for large black hole masses, is associated with star formation for the majority of cases in dwarf galaxies \citep{baldassare2016}, further emphasizing the limitations of optical studies in finding AGNs in low mass bulge mass regime. 

What is the path forward in identifying AGNs in the low bulge mass regime?  Based on the hunt for AGNs in the low bulge mass regime thus far, it is clear that multi wavelength observations are crucial in our attempt to arrive at a complete census, and that studies based on only one technique will be incomplete.  Given that we now know that both obscuration and host galaxy contamination is often both at play in the low luminosity regime, a promising prospect will be sensitive mid-infrared spectroscopic observations of the [NeV] emission line that will be enabled by the {\it James Webb Space Telescope}, which can robustly separate starbursts from AGNs \citep{abel2008} and potentially yield as much as 4 times as many AGNs in galaxies that lack classical bulges compared to optical studies \citep{satyapal2008}.

%{\textcolor{red}{elaborate more here}}

% things to add: 1) mention that apart from scatter in Lehmer relation, there are significant uncertainties in the estimate of the SFR, which ranges from optical recombination line measurements, SED fitting, and IRAS derived values, all of which can yield significantly different results (find reference for this). 2) Emphasize that the observations are not definitive by any means. 3) Calculate eddington mass and compare to J1329 and other galaxies in this class.  Comment on M-sigma in the low mass regime. 4) Comment on future prospects. Maybe JWST and NeV?5) Add comments to Jess' paper and impact on seeds.5) Point out shortcomings of all methods - including near-IR not good at uncovering AGNs even in known optically identified or hard X-ray selected AGNs (Koss et al in prep).

\section{Summary and Conclusions}

We have conducted the second follow-up X-ray observation of a newly discovered population of  bulgleless galaxies that display extremely red mid-infrared colors highly suggestive of a dominant AGN despite having no optical signatures of accretion activity. Our main results can be summarized as follows: \\

\begin{enumerate}
\item{Using \textit{XMM-Newton} observations, we have confirmed the presence of an unresolved X-ray source coincident with the nucleus of \gal, an optically normal bulgeless galaxy with red infrared colors obtained from \textit{WISE}.\\}

\item{The observed X-ray luminosity (uncorrected for intrinsic absorption) of \gal\ is   $L_\mathrm{2-10~keV}=(1.1\pm0.4)\times10^{40}$~erg~s$^{-1}$ , which is a factor of 3-4 lower than the luminosity expected from XRBs in the host galaxy. But given the scatter in the relation, the observed X-ray luminosity is not significantly above the relation.\\}

\item{While the X-ray observations alone do not require the presence of an AGN, a multiwavelength investigation of the X-ray, near-IR, and mid-IR activity of \gal{} is consistent with the presence of a highly obscured and energetically dominant AGN with column density $N_\mathrm{H} > 10^{24}$~cm$^{-2}$.\\}

\item{The hard X-ray luminosity of the putative AGN corrected for absorption is $L_\mathrm{2-10~keV}\sim3\times10^{42}$~erg~s$^{-1}$, which, depending on the bolometric correction factor, corresponds to a bolometric luminosity of the AGN of $L_\textrm{bol.}\sim 6\times10^{43}$~erg~s$^{-1}$-$3\times10^{44}$~erg~s$^{-1}$, and a lower mass limit for the black hole of  $M_\textrm{BH}\simeq2\times10^6$~M$_\sun$ - $2\times10^7$~M$_\sun$, based on the Eddington limit.\\}
\item{There is no evidence for coronal lines or broad lines  in the near-infrared spectrum of ~\gal\, suggesting that the putative AGN is obscured even in the K band.\\}
 
\item{These observations demonstrate the ambiguity associated with identifying AGNs in the low luminosity regime and highlight the need for a careful mutliwavelength investigation.\\}

\end{enumerate}

The X-ray observations of ~\gal\  presented here are one of the two only follow-up observations of  bulgeless galaxies with red mid-infrared colors discovered by {\it WISE}.  Both targets point to the presence of optically obscured AGNs, suggesting that some low mass and bulgeless galaxies do harbor AGNs. These observations may suggest that optically normal AGNs in bulgeless galaxies are more common than previously thought, and emphasize the need for more follow-up multi wavelength studies of this population.

\section{Acknowledgements}

We gratefully acknowledge the anonymous referee for a very thorough and insightful review that improved this manuscript.  We also gratefully acknowledge very helpful discussions with Daniel Stern on the ambiguities associated with mid-infrared color selection in low mass and bulgeless galaxies. The plots in this paper greatly benefited from previously written scripts by Paul McNulty, with whom it has been a joy to work. N.\init J.\init S.~and S.\init S.~gratefully acknowledge support by the \textit{XMM} Guest Investigator Program under NASA Grant NNX14AF02G  and a Mason 4-VA Innovation grant.  R.C.H. acknowledges support from an Alfred P. Sloan Research Fellowship and from the National Science Foundation through grant number 1515364. J.\init L.\init R.\init acknowledges support from NSF-AST 000167932.  This research has made use of the NASA/IPAC Extragalactic Database (NED) which is operated by the Jet Propulsion Laboratory, California Institute of Technology, under contract with the National Auronautics and Space Administration. We also gratefully acknowledge the use of the software TOPCAT \citep{Taylor2005} and Astropy \citep{astropy2013}.\\

\end{document}